\documentclass[aps,prb,twocolumn,preprintnumbers,showpacs,showkeys]{revtex4}
\usepackage{graphicx}
\usepackage{amsfonts}
\usepackage{amsmath}
\usepackage{amssymb}
\usepackage{amsfonts}
\usepackage{graphicx}

\begin{document}
\bibliographystyle{unsrt}
\title[Injection]{Current induced spin flip scattering at interfaces in
noncollinear magnetic multilayers}
\author{Peter M Levy}
\affiliation{Department of Physics, 4 Washington Place, New York University, New York,
New York 10003}
\author{Jianwei Zhang}
\affiliation{Department of Physics, 4 Washington Place, New York
University, New York, New York 10003} 

\begin{abstract}
We show that when one drives a charge current across noncollinear magnetic
layers the two electron Coulomb scattering creates a spin flip potential at
interfaces. This scattering is found when the interface potential is updated
due to the spin accumulation attendant to charge flow, and it contributes in
linear response to the current. With this scattering there is an injection
of transverse spin distributions in a layer that propagate, and that in the
steady state lead to spin currents transverse to the magnetization.
\end{abstract}

\volumeyear{year}
\volumenumber{number}
\issuenumber{number}
\eid{identifier}
\date[Date:]{\today}
\received[Received text]{date}

\maketitle
Our current knowledge of the electronic structure of
magnetic multilayers is based on effective single electron
approximations. In nonmagnetic (normal) layers states are spin
degenerate while in the magnetic they are spin split; therefore
one can construct a coherent combination of spin up and down
states in the normal layers, while in the magnetic this is not
possible because there is no coherence (correlations) between
states in spin split bands.\cite{stiles} This loss of spin
coherence as one crosses a normal/ferromagnetic layer (N/F)
interface has been identified as the origin for the inability of
injecting transverse spin currents from a normal metal into a 3d
transition-metal ferromagnet.\cite{discontinuities} The studies to
date of the coherence transmitted by interface scattering have
concluded that spin currents across this interface can induce only
a transverse coherence between electrons on different sheets of
the spin split Fermi surface of the
ferromagnet.\cite{discontinuities} Here we revisit this question
and find a current driven correction to the transmission
amplitudes coming from the two electron Coulomb interaction at
interfaces that allows the transverse spin coherence to be carried
by the electron distribution function for each spin split band. As
we show while this is an out of equilibrium effect, its inclusion
in transport calculations enters in linear response. In another
publication we show that this current induced coherence make it
possible for spin currents to be continuous across the interface
in the \textit{steady state}.\cite{jianwei} Inasmuch as one
includes current driven accumulations in the bulk of the layers
when calculating the steady state currents we conclude that to
discuss spin currents in noncollinear magnetic structures it is
necessary to self consistently update the transmission and
reflection coefficients due to the presence of out of equilibrium
spin distributions at interfaces.

To relate distribution functions $f_{ss^{\prime }}(k,r)$ across an
interface one resorts to the transfer matrix \cite{bauer}
\begin{align}
f_{ss^{\prime }}^{>}(k,0^{+})&=\sum_{mm^{\prime }\hat{k}^{\prime
}}T_{mm^{\prime }\text{ }ss^{\prime }}f_{mm^{\prime }}^{>}(\hat{k}^{\prime
},\varepsilon _{k},0^{-}) \notag \\
&+\sum_{mm^{\prime }\hat{k}^{\prime \prime
}}R_{mm^{\prime }\text{ }ss^{\prime }}f_{mm^{\prime }}^{<}(\hat{k}^{\prime
\prime },\varepsilon _{k},0^{+}),  \label{1}
\end{align}%
where the transmission and reflection coefficients (probabilities) are
\begin{align}
T_{mm^{\prime }\Rightarrow ss^{\prime }}& =T_{mm^{\prime }ss^{\prime }}=t_{m%
\text{ }s}\ast t_{m^{\prime }s^{\prime }}^{\ast },  \notag \\
R_{mm^{\prime }\Rightarrow ss^{\prime }}& =R_{mm^{\prime }ss^{\prime }}=r_{m%
\text{ }s}\ast r_{m^{\prime }s^{\prime }}^{\ast },  \label{2}
\end{align}%
and $t_{ms}(k^{\prime },k,\varepsilon _{k}),r_{ms}(k^{\prime \prime
},k,\varepsilon _{k})$ are the\ transmission and reflection amplitudes for
elastic scattering\textit{.} At the interface between a normal and
ferromagnetic layer the momenta $k,k^{\prime },k^{\prime \prime }$\ are
either $k_{M}$ or $k_{m}$, for a majority and minority bands of the
ferromagnetic layer and $k_{n}$ for a normal metal layer. We make the
"isotropy" assumption that allows one to write one coefficient for all
momentum directions, albeit one should still keep a band index, i.e., $%
T_{mm^{\prime }\Rightarrow ss^{\prime }}(k^{\prime },k,\varepsilon
_{k})\thicksim T_{mm^{\prime }\Rightarrow ss^{\prime }}(\varepsilon _{k})$.%
\cite{bauer} In a spin polarized single electron description the spin split
bands of the magnetic layer have only one spin index, $s=s^{\prime }$, for
each momentum $k$ ,i.e., $f_{ss^{\prime }}=f_{s}$, however when one is
propagating currents across noncollinear structures one has to envisage the
possibility of off diagonal components of the distribution function $%
f_{ss^{\prime }}^{>}(k,0^{+})$ with $s\neq s^{\prime }$ within a screening
length about the interface. This represents a coherence between states in
different bands; it does not represent an admixture of spin within the band
although the scattering we find is capable of this.\cite{coherence} The
indices $m,m^{\prime }$ refer to the spin of the electron in the unsplit
Fermi sea of the normal metal layer; in nonmagnetic layers it is possible to
construct coherent combinations of up and down spin states $m\neq m^{\prime
}.$ Scattering at interfaces in metallic structures is localized within a
screening length about the interface and represented by amplitudes $%
t_{m_{l}\rightarrow m_{r}}$,$r_{m_{l}\rightarrow m_{l}^{\prime }}$ ; the
first describe the transmission from, say, the left layer to the right, the
second the reflection back into the left layer, and the $m^{\prime }s$ are
the components of the electron's spin.

The scattering potential, $t$ matrix, for conduction electrons at an
interface between two metals has contributions from differences in, the
periodic background potentials, the kinetic energy and the Coulomb
interaction between the electrons; when written as an effective one electron
potential only the latter changes due to current driven spin accumulation on
the Fermi surface.\ To find these corrections we focus on the
electron-electron contribution to the $t$ matrix. The one electron
scattering amplitudes at interfaces are derived form the two electron
Coulomb interaction
\begin{equation}
V_{coulomb}=\frac{1}{2}{\displaystyle\sum\limits_{\substack{ k_{1}..k_{4}
\\ s,s^{\prime }}}}V(k_{1s}k_{2s^{\prime }}k_{3s^{\prime
}}k_{4s})c_{k_{1}s}^{\dagger }c_{k_{2}s^{\prime }}^{\dagger
}c_{k_{3}s^{\prime }}c_{k_{4}s},  \label{3a}
\end{equation}%
where $V(k_{1s}..k_{4s})$ are the matrix elements of the Coulomb interaction
between states on either side of the interface. For our purposes the range
of integration is limited to the screening length about the interface. As we
are working at the interface between dissimilar well screened metals the
translational invariance of the background is broken; therefore we do not
reduce the four to three momenta. To determine the one electron scattering
amplitude, $t_{m_{l}\rightarrow m_{r}}$,$r_{m_{l}\rightarrow m_{l}^{\prime }}
$, that arises from this Coulomb interaction we write the Coulomb
interaction between electrons in the one electron states in the bulk of the
layers on either side of the interface, and to reduce it to a one electron
operator we take the expectation value over a pair of annihilation and
creation operators in Eq.(\ref{3a}). In the lowest order distorted wave Born
approximation the spin dependent part of the correction to the transition
matrix between one electron states on the two sides of the interface is \cite{messiah}
\begin{equation}
t_{m\rightarrow s}=\left\langle k_{2},s\left\vert t_{op}\right\vert
k_{n},m\right\rangle ,  \label{4a}
\end{equation}%
where
\begin{equation}
t_{op}\equiv -\frac{1}{2}g({\varepsilon }_{k_{2}}){\displaystyle%
\sum\limits_{k_{1}k_{3}}}\left\langle c_{k_{1}m}^{\dagger
}c_{k_{3}s}\right\rangle V(k_{1m}k_{2s}k_{3s}k_{nm})c_{k_{2}s}^{\dagger
}c_{k_{n}m},  \label{5a}
\end{equation}%
$|k_{n},m>$\ \ refers to states on the Fermi surface of the normal layer,
and $<k_{2},s|$ to those on the ferromagnetic spin split Fermi surface, so
that $k_{2}=k_{M}$ and $k_{m}$. For elastic scattering $\delta (\varepsilon
_{k_{n}}-\varepsilon _{k_{2}})$ the density of states $g({\varepsilon }%
_{k_{2}})$ enters when one averages the amplitude $t_{m\rightarrow s}$ over
the energy $\varepsilon _{k_{n}}$. While the states in the ferromagnetic
layers are pure spin states, e.g., only an up spin goes in the majority
sheet, we are looking for current induced coherences in the distribution
function, a statistical density matrix, between states of opposite spin;
therefore we do not immediately associate a spin $s$ to a state $k_{2}$,
e.g., we will be looking for matrix elements for both spin directions on
each sheet of the Fermi surface of the ferromagnetic layer that are induced
when the system is out of equilibrium.

The expectation value $\left\langle c_{k_{1}m}^{\dagger
}c_{k_{3}s}\right\rangle $ is between states on opposite sides of the
interface and is found by self consistently evaluating it there; as we will
be interested only in elastic scattering the energy of the states entering
this expectation value are the same, $\varepsilon _{k_{1}}=\varepsilon
_{k_{3}}$, but their directions in $k$ space can be different. At the
interface between normal and ferromagnetic layers (N/F) the\ scattering
potential, while spin dependent, is diagonal in spin space, i.e., there is a
unique spin direction, and there are no elements $m\neq s$ when the system
is in \textit{equilibrium}, and $\left\langle c_{k_{1}m}^{\dagger
}c_{k_{3}s}\right\rangle \thicksim n_{s}^{int}\delta _{sm}$ ; it also
follows that $t_{ms}^{eq}=t_{s}\delta _{sm}$. When driving current across a
magnetic multilayer spin accumulates on the Fermi surface of the normal
layers so as to support a spin current across them in steady state,
therefore in the presence of a spin current the spin polarized distribution,
$\delta n_{s}^{l},$ of the Fermi surface of the magnetic layer $\hat{M}_{l}$
upstream from the N/F$_{r}$ interface is superimposed on the Fermi surface
of the normal layer.

For \textit{noncollinear} magnetic layers the magnitude of the current
driven accumulation \textit{in the interfacial region} N/F$_{r}$ coming from
the left magnetic layer, $\delta n_{s}^{int},$ is uncertain; this can only
be ascertained by a calculation of the transmission amplitudes at N/F$_{r}$
when one superimposes $\delta n_{s}^{l}$ on the Fermi surface (at the Fermi
energy) of the normal layer.\cite{ingrid} In linear response we do know it
is quantized along the magnetization $\hat{M}_{l}$ of the magnetic layer
upstream while the operators entering the expectation value in the
interfacial region ( see Eq.(\ref{5a})) are quantized along $\hat{M}_{r}$
the magnetization of the ferromagnet at the N/F$_{r}$ interface. To
determine the additional contribution from $\delta n_{s}^{l}$ to the
transmission amplitude, Eq.(\ref{5a}), it is necessary to rotate the
accumulation referred to $\hat{M}_{l}$ to states quantized along $\hat{M}_{r}$; we find%
\begin{equation}
\delta t_{op}\equiv A_{ms}(k_{2},k_{n})c_{k_{2}s}^{\dagger }c_{k_{n}m},
\label{6a}
\end{equation}%
where%
\begin{eqnarray}
& A_{ms}(k_{2},k_{n})=V_{s}(k_{s}k_{n};\varepsilon _{k_{2}}) \notag \\
&\times\left\{
\begin{array}{c}
\left[ \delta n_{s}^{int}\cos ^{2}\theta /2+\delta n_{-s}^{int}\sin
^{2}\theta /2\right] \delta _{sm} \\
-\frac{i}{2}\left[ \delta n_{s}^{int}-\delta n_{-s}^{int}\right] \sin \theta
\delta _{s,-m}%
\end{array}%
\right\} ,  \label{7a}
\end{eqnarray}%
\begin{equation}
V_{s}(k_{2},k_{n};\varepsilon _{k_{2}})=-%
{\frac{1}{2}}%
g_{s}({\varepsilon }_{k_{2}})\left\langle \left\langle
V(k_{2s},k_{nm};\varepsilon _{_{F}})\right\rangle \right\rangle ,  \label{8a}
\end{equation}%
\begin{eqnarray}
&\left\langle \left\langle V_{s}(k_{2s},k_{nm};\varepsilon )\right\rangle
\right\rangle =  \notag \\
&\frac{1}{(4\pi)^2 } \! \int\limits_{\varepsilon
_{k_{1}}=\varepsilon _{k_{3}} =\varepsilon } \! \! \! \! \! \! \! d\Omega
_{k_{1}} \! \int\limits_{\varepsilon _{k_{3}}} \! d\Omega _{k_{3}}V_{s}(\hat{k}%
_{1},k_{2s},\hat{k}_{3},k_{nm}),  \label{9a}
\end{eqnarray}%
Here $\delta n_{s}^{int}$ is the occupancy of the spin states on the Fermi
surface at the interface which have been polarized by the upstream magnetic
layer $\hat{M}_{l}$.

The out of equilibrium accumulation and current transmitted across an
interface N/F by the equilibrium $T$ matrix is found from Eq.\ (\ref{1}) by
integrating over all $k$ states. For the diagonal spin components $%
f_{ss^{\prime }}^{>}=f_{s}^{>}$ $\delta _{ss^{\prime }}$ the out of
equilibrium distribution only exists on the Fermi surface, because $%
f_{mm^{\prime }}^{>}(k^{\prime },\varepsilon _{k},0^{-})\thicksim \delta
f_{m}(\hat{k}^{\prime })\delta _{mm^{\prime }}\delta (\varepsilon
_{k}-\varepsilon _{F})$. For spin transport across \textit{noncollinear}
magnetic layers the state $|k_{n},m>$ is quantized along the magnetization $%
\hat{M}_{l}$ of the magnetic layer upstream while the operators in the $t$
matrix are quantized along $\hat{M}_{r}$ the magnetization of the
ferromagnet at the N/F interface, i.e., $\theta =\cos ^{-1}($ $\hat{M}%
_{l}\cdot \hat{M}_{r})$, therefore it is necessary write $|k_{n},m>$ in
terms of $|k_{2},s>$. By rotating this state we find the average of the
transmission coefficient in equilibrium is
\begin{equation}
T_{mm\Rightarrow ss}(\varepsilon _{F})=\left\vert t_{s}\right\vert
^{2}\left\{ \cos ^{2}\theta /2\delta _{s^{{}}m}+\sin ^{2}\theta /2\delta
_{s^{{}}-m}\right\} ,  \label{10a}
\end{equation}%
where $t_{s}=t_{ms}^{eq}\delta _{sm}$. As the spin diagonal term $\thicksim
\delta _{sm}$ in Eq.(\ref{7a}) is proportional to the accumulation or
current it does not enter in linear response in the $T$ matrix that connects
the spin diagonal part of out of equilibrium distribution functions, $%
f_{s}^{>}$ and $\delta f_{m}(\hat{k}^{\prime })$, as they themselves produce
the accumulation and current when integrated over the Fermi surface. However
the off diagonal (spin flip) term$\thicksim \delta _{s^{{}}-m}$ in Eq.(\ref%
{7a}) is uncompensated because the accumulation coming from the right layer
at the N/F$_{r}$ is diagonal in spin space, and cannot offset an off
diagonal term coming \ from the left. At the other interface of the normal
layer, N/F$_{l}$ , the accumulation arising from the right layer $\delta
n_{s}^{r}$ cannot be fully compensated from that arising from the left layer
and we have an uncompensated contribution $\delta t_{op}\thicksim \delta
_{s,-m}$. \cite{stiles1}

The current driven coherences $f_{ss^{\prime }}^{>}$ with $s\neq s^{\prime }$
do not represent populations, that in equilibrium are given by the Fermi
Dirac distribution function, therefore their contribution to the
transmission across an interface is \textit{not} limited to the Fermi
surface ($\varepsilon _{k}\neq \varepsilon _{F}$) and has contributions from
the entire Fermi sea. As the scattering potential, Eq.(\ref{6a}), is
proportional to the accumulation or current its contribution in linear
response to the spin coherence transmitted across an interface is found by
using the equilibrium distribution in the normal layer $f^{0}(\varepsilon
_{k},0^{-})\delta _{mm^{\prime }}$ in Eq.(\ref{1}), and by using an
equilibrium transmission amplitude to find the update to the equilibrium $T$
matrix in the presence of a current. As the distribution function entering
the right hand side of Eq.(\ref{1}) does not depend on the spin index $m,$
we find, in linear response, the contribution of the spin flip transmission
amplitude Eq.(\ref{6a}) to the off diagonal transmission coefficient $%
\sum_{m}\delta T_{mm\Rightarrow ss^{\prime }}$ for states on the Fermi
surface comes from integrating over the Fermi sea
\begin{eqnarray}
& f_{ss^{\prime }}^{>}(\hat{k},\varepsilon _{F},0^{+})  \notag \\
&=\sum_{k,m}\left\{
t_{ms}^{eq}\ast \delta t_{ms^{\prime }}^{\ast }+\delta t_{ms}\ast
t_{ms^{\prime }}^{eq\ast }\right\} f^{0}(\varepsilon _{k},0^{-})  \notag \\
&=\delta n_{z}^{int}\sin \theta \left\{ \Re \overline{t_{s}V_{s}^{\ast
}}\left[ \sigma _{y_{r}}\right] _{ss^{\prime }}+\Im \overline{%
t_{s}V_{s}^{\ast }}\left[ \sigma _{x_{r}}\right] _{ss^{\prime }}\right\} ,
\label{13a}
\end{eqnarray}%
where%
\begin{equation}
\overline{t_{s}V_{s}^{\ast }}=\int d\varepsilon _{k}f^{0}(\varepsilon
_{k})g_{n}(\varepsilon _{k})\left\langle \left\langle
t_{s}(k_{2},k_{n};\varepsilon _{k})V_{s}^{\ast }(k_{s}k_{n};\varepsilon
_{k})\right\rangle \right\rangle ,  \label{13b}
\end{equation}%
$t_{s}(k_{2},k_{n};\varepsilon _{k})$ is the equilibrium transmission
amplitude, and the angular brackets are defined in Eq.(\ref{9a}). Here the
out of equilibrium densities are $\delta n_{z}\equiv
{\frac{1}{2}}\left[ \delta n_{\uparrow }-\delta n_{\downarrow }\right]$, and to arrive
at this result we made the isotropy assumption for the transmission
coefficients so that the angular averages over these coefficients are done
independently of those over the distribution function.

This transmission coefficient is current driven so that in linear response
it acts on the equilibrium distribution function at the Fermi surface in Eq.(%
\ref{1}) to produce an out of equilibrium distribution $f_{ss^{\prime }}^{>}$
; whereas the transmission Eq.( \ref{10a}) acts on the out of equilibrium
distribution on the Fermi surface. By writing $\delta n_{z}^{int}\equiv
\alpha \delta n_{z}^{l}$ and%
\begin{equation}
\delta n_{z}^{l}=%
{\frac{1}{2}}
\sum\limits_{m}\left[ \sigma _{z_{l}} \right] _{mm}\int d\Omega _{k^{\prime
}}\delta f_{m}(\hat{k}^{\prime }),  \label{14a}
\end{equation}%
we find the two contributions, Eqs.(\ref{10a}) and (\ref{13a}), to the $T$
matrix can be written as
\begin{align}
&f_{ss^{\prime }}^{>}(\hat{k},\varepsilon _{F},0^{+})=\sum_{m}
T_{mm\Rightarrow ss}(\varepsilon _{F})\delta _{ss^{\prime }} \delta f_{m}(\hat{k}^{\prime })
+\sum_{m}{\frac{1}{2}}\alpha \sin \theta \notag \\
& \times \left[ \sigma _{z_{l}}\right] _{mm}\left\{ \Re %
\overline{t_{s}V_{s}^{\ast }}\left[ \sigma _{y_{r}}\right] _{ss^{\prime }}+%
\Im \overline{t_{s}V_{s}^{\ast }}\left[ \sigma _{x_{r}}\right]
_{ss^{\prime }}\right\}
 \delta f_{m}(\hat{k}^{\prime }).  \label{14b}
\end{align}%
Whereas the first term is the equilibrium scattering at the interface in the
conventional approach, the second depends on the spin accumulation or
current at the interface which is proportional to $\delta f_{m}(\hat{k}%
^{\prime })$, and produces an off diagonal component of the spinor density
matrix at the interface. While $V_{s}$ is a sum over the Fermi surface ( see
Eq.(\ref{9a}) with $\varepsilon =\varepsilon _{F}$) and $t_{s}$ over the
entire Fermi sea their difference is made up in the integration over the
equilibrium distribution function that enters Eq.(\ref{13b}) which is over
Fermi sea. We can only do a meaningful comparison, of the transmission
amplitudes in equilibrium and another when one superimposes $\delta n_{z}^{l}
$ on the Fermi surface of the normal layer, by calculating the interface
scattering for these two situations; interalia this will determine the
constant $\alpha $ which indicates how much of the spin accumulation $\delta
n_{z}^{l}$ penetrates into the interfacial region so as to affect the
interface scattering potential.\cite{ingrid}

By solving the Boltzmann equations of motion we have found that the
magnitude of the "out of equilibrium" $\delta T$ relative to $T$ controls
the amount of transverse spin accumulation, but its very existence
guarantees the continuity of spin currents at the interface when a steady
state is achieved, i.e., to eventually achieve a continuous spin current in
the steady state it is necessary that the interface scattering potential has
matrix elements that inject transverse spin distributions into a magnetic
layer with well defined momentum so that they can propagate past the
interface, and into the bulk of a layer.\cite{jianwei} Without this
additional scattering at the interface we have found a discontinuity in the
spin current at all times; while this is found by setting the explicit time
derivative of the distribution function, $\partial _{t}f(k,r,t)$, to zero
this solution differs from steady state inasmuch as there is a constant loss
of transverse spin current at the interface, i.e., it is questionable
whether one can call this a true steady state. The transverse spin current $%
j_{y}$ and $j_{x}$(this exists if the transmission amplitudes are complex)
we calculate in the ferromagnetic layer, by using Eq.(\ref{13a}) or (\ref%
{14b}), arises from injecting the transverse component of the incoming spin
current from a normal layer in such a manner that it excites a transverse
mode of propagation in the bulk of the ferromagnetic layer. Our scattering
potential provides for the transfer of the transverse spin distribution
across an interface in such a manner that it can propagate past the
interfacial region and into the bulk of a magnetic layer; it do not suffer
from the dephasing previously found when one considered the transverse spin
current that arises from scattering onto different sheets of the Fermi
surface, e.g.,
\begin{equation}
T_{mm\Rightarrow ss^{\prime }}\thicksim \lbrack t_{s}(k_{M})t_{s^{\prime
}}^{\ast }(k_{m})]\left\{ i/2\sin \theta \delta _{s^{\prime }-s}\right\} ,
\label{16a}
\end{equation}%
where $k_{M/m}$ represent the Fermi momenta of the majority/minority bands.
Among other things the distribution function \ $f_{ss^{\prime }}^{>}(\hat{k}%
,\varepsilon _{F},0^{+})$ found by using this $T$ in Eq.(\ref{1}) does not
have a unique velocity $v(k),$as $k_{M}\neq $ $k_{m}$, therefore it cannot
be used to calculate a current within the Boltzmann approach. The existence
of this scattering between bands limits transverse currents to a region 1-2
monolayers of the interface,\cite{discontinuities} but it does not negate
the current induced interface scattering mechanisms, Eq.(\ref{13a}), that
allow for injection into a coherent mode of transverse spin propagation on
another length scale.

The spin current transmitted by the conventional coefficients, Eq. ( \ref%
{10a}), is parallel to the local magnetization; it does not have a
transverse component. If the interface scattering Eq.(\ref{1}) is unable to
scatter electrons into transverse spin distribution functions $f_{ss^{\prime
}}(k)$ $s^{\prime }\neq s$ which can propagate past the interfacial region
when the current is first turned on, the transverse component of the spin
current never leaves the interfacial region; not even in steady state. In a
manner of speaking the transverse spin accumulation will remain confined to
the interface and one can talk about the spin current as being discontinuous
even in the steady state. The two electron scattering across an interface
Eq.(\ref{3a}) creates the scattering potential, Eqs.(\ref{6a}) and (\ref{13a}%
) that allows for a coherent transmission of spin information at the
interface. However this only exists at the interface, within a screening
length of the interface; without a current this electron distribution does
not go beyond the interface scattering region. In the presence of a current
once created at the interface the transverse distribution propagates
according to the Boltzmann equation; as the distribution $f_{ss^{\prime }}$
for $s^{\prime }\neq s$ does not commute with the Hamiltonian in a magnetic
layer ( which is described in the spin polarized single electron
approximation) we show in another publication this leads to distributions
that precess, and that it is the exchange splitting in this approximation $%
J(k)$ that controls this precession.\cite{jianwei} Eventually,this leads to
spin accumulation transverse to the magnetization of the magnetic layer, so
that in steady state the spin current is continuous across the N/F interface.%
\cite{jianwei}

In conclusion our scattering potential provides for the transfer of the
transverse spin distribution across an interface in such a manner that it
can propagate past the interfacial region and into the bulk of a magnetic
layer. The spin flip scattering potential Eq.(\ref{6a}) lies outside the
conventional spin polarized single electron treatment of the scattering at a
N/F interface as one does not usually envisage when calculating the
transmission amplitude a distribution from the spin polarized Fermi surface
of a neighboring magnetic layer $\hat{M}_{l}$ being superimposed onto the
normal layer.\cite{bauer1} However when one uses an approximate conductivity
that is local (short ranged), e.g., when using only the bubble conductivity
in the Kubo formalism or the Boltzmann layer by layer approach, in order to
account for the long range nature of the conductivity we are required to
posit this in the presence of an electric field. \cite{our work} \cite{kane}
The strength of the spin flip potential relative to the ordinary scattering
at the interface can be ascertained from calculations of the transmission
amplitudes in the presence of spin accumulation in the normal layer.

We would like to thank Arne Brataas, Andrew Kent, Ingrid Mertig, and Shufeng
Zhang for very helpful discussions. This work was supported by the National
Science Foundation (Grant DMR 0131883).


\begin{thebibliography}{99}
\bibitem{coherence} Coherences in the distribution function are the off
diagonal elements of statistical density matrix; here we are interested in
those induced by currents across noncollinear magnetic structures at the
Fermi surface.
\bibitem{bauer1} The possibility for spin flip scattering at interfaces was
envisaged in the papers of Bauer's group, see Refs.2 and 4, but their origin
was not discussed nor was their role included in the analyses.
\end{thebibliography}
\end{document}